\documentclass[aps,prd,showpacs,groupedaddress]{revtex4}

\usepackage{amsmath,amssymb,bm}

\usepackage{graphicx}
\usepackage{amsfonts}
\usepackage{amssymb}
\usepackage{amsbsy}
\usepackage{amsmath}
\usepackage{latexsym}
\usepackage{bm}
\usepackage{color}
\usepackage{comment}

\def\p {\partial}
\def\t {\tilde}
\def\be {\begin{equation}}
\def\ee  {\end{equation}}
\def\bea {\begin{eqnarray}}
\def\eea {\end{eqnarray}}
\def\nn {\nonumber}
\def\sgn {\mbox{sgn}}

\begin{document}

\title{3D gravity with dust: classical and quantum theory}

\author{Viqar Husain}
\email[]{vhusain@unb.ca}

\author{Jonathan Ziprick}
\email[]{jziprick@unb.ca}

\affiliation{University of New Brunswick\\
Department of Mathematics and Statistics\\
Fredericton, NB E3B 5A3, Canada}

\date{\today}

\begin{abstract}
    
We study the Einstein gravity and dust system in three spacetime dimensions as an example of a non-perturbative quantum gravity model with local degrees of freedom. We derive the Hamiltonian theory in the dust time gauge and show that it has a rich class of exact solutions. These include the Ba\~nados-Teitelboim-Zanelli black hole, static solutions with naked singularities and travelling wave solutions with dynamical horizons. We give a complete quantization of the wave sector of the theory, including a definition of a self-adjoint spacetime metric operator. This operator is used to demonstrate  the quantization of deficit angle and the fluctuation of dynamical horizons.

\end{abstract}

\pacs{04.20.Fy, 04.20.Jb, 04.60.Ds}

\maketitle

\section{Introduction}

The difficulty in formulating a four dimensional theory of quantum gravity has led to the study of many simpler models. These include symmetry reductions of 4D general relativity \cite{FernandoBarbero:2010qy} and dimensional reduction to 3D gravity \cite{Deser:1983tn,Ashtekar:1989qd,Moncrief:1990mk,Carlip:1994ap,Carlip:1998uc,Witten:1988}. There is a large volume of literature on the latter, which includes pure gravity with point defects and/or topological degrees of freedom \cite{tHooft:1993cl,tHooft:1993qu,Freidel:2004,Ziprick:2015}, topologically massive gravity \cite{Carlip:2008eq}, and higher spin gravity. While some of these (lower dimensional) simplifications have allowed for covariant quantization, there is relatively little work on the canonical quantization of any gravity-matter model. 

Our purpose in this paper is to develop a 3D model of gravity with matter which has the potential for complete quantization. This would reveal insights into quantum gravity in a setting with local degrees of freedom. The pressureless dust matter we use is perhaps the simplest such model, but it is sufficiently non-trivial in that there is a rich class of classical solutions, including ones with dynamical horizons. Such solutions are of much interest at the quantum level: questions such as what is the quantum analog of a classical dynamical horizon remain unanswered, and are key to understanding what is a ``quantum black hole'' or  a ``quantum trapped surface"  \cite{Husain:2004yy}.

With these issues in mind we begin by formulating a canonical theory of 3D gravity coupled to pressureless dust. This is a special case of the Brown-Kuchar \cite{Brown:1994py} model which is designed to provide a dynamical matter reference system for gravity in 3+1 dimensions.  It was used to  give a physical Hilbert space setting for loop quantum gravity in the dust time gauge \cite{Husain:2011tk, Husain:2011tm, Husain:2013yya}, and added as an additional world sheet field  in the bosonic string to yield a curious extension of that theory \cite{Gegenberg:2012pg}. 

We will see that in the 2+1  model, the dust time gauge gives a physical Hamiltonian that describes  the dynamics of one local geometry degree of freedom; this remains in the circularly symmetric setting we consider in detail. The model  also provides an example of the transfer of a matter degree of freedom to a geometric one; this may provide a useful viewpoint for quantum gravity in a more general setting, distinct from the strict conventional separation of matter and geometry degrees of freedom. In Section II we develop the circular-symmetry-reduced theory and in Sec. III we give the gauge fixed theory. In Section IV we give several types of classical solutions, followed by the construction of a quantum theory of the system in Section V, with focus on the travelling wave solutions. The concluding section is a summary of our results and a discussion of further questions.

\section{Hamiltonian theory}
In units where $8 G = c = 1$ the action for gravity and dust is a sum of the two components
\be
\label{action}
S = S_G + S_D.
\ee
Let us  consider this action  defined on  a three-dimensional manifold with topology $\Sigma \times \mathbb{R}$. The gravitational part of the action is
\be
S_G = \frac{1}{2\pi}\int dx^3 \sqrt{g}(^{(3)}R - 2\Lambda),
\ee
where $^{(3)}R$ is the scalar curvature of spacetime and $\Lambda$ is the cosmological constant.
The dust action is 
\be
\label{dust_action}
S_D = - \frac{1}{4 \pi} \int dx^3 \sqrt{g} m \left(g^{\mu \nu} \p_\mu \phi \p_\nu \phi + 1 \right),
\ee
where $m(x)$ is a function of the spacetime coordinates.

 To derive the   Hamiltonian formulation we use   the ADM (Arnowitt-Deser-Misner) parametrization of the  line element
\be
ds^2 = -N^2 dt^2 + q_{ab} (dx^a + N^a dt)(dx^b + N^b dt),
\ee
where $q_{ab}$ is the space metric, $N$ is the lapse function and $N^a$ is the shift vector. With this the  gravitational part of the action takes the well-known canonical form (see eg. \cite{Carlip:1998uc})
\be
S_G = \frac{1}{2\pi} \int dx^3 \left(\tilde{\pi}^{ab} \dot{q}_{ab} - N \mathcal{H}^G - N^a \mathcal{C}^G_a \right),
\ee
where $\tilde{\pi}^{ab}$ is the (density weight one) momentum  conjugate to $q_{ab}$.  $N$ and $N^a$ appear as the Lagrange multipliers corresponding respectively to the Hamiltonian and diffeomorphism constraints 
\bea
\mathcal{H}^G &=& \sqrt{q} \left(-^{(2)}R + \frac{1}{q} ( \tilde{\pi}^{ab} \tilde{\pi}_{ab} - \tilde{\pi}^2 ) + 2 \Lambda \right) , \\
\mathcal{C}^G_a &=& -2 \nabla_a \tilde{\pi}^a_b,
\eea
where  $q \equiv \det q_{ab}$, $\tilde{\pi} \equiv \tilde{\pi}^a_a$ and  $^{(2)}R$ is the Ricci scalar of the spatial hypersurface.

 The canonical dust action  is obtained starting with the momentum 
 \be
\label{P-phi}
P_\phi :=  \frac{\delta S_D}{\delta \dot{\phi}} = \frac{\sqrt{q} m}{N} \left( \dot{\phi} - N^a \p_a \phi \right),
\ee
which leads to 
\bea
S_D &=& \frac{1}{2\pi} \int dx^3 \left( P_\phi \dot{\phi} - N \mathcal{H}^D - N^a \mathcal{C}^D_a \right) , \\
\mathcal{H}^D &=& \frac{1}{2} \left(\frac{P_\phi^2}{m\sqrt{q}} + m \sqrt{q} (q^{ab}\p_a \phi \p_b \phi +1 ) \right), \\
\mathcal{C}^D_a &=& P_\phi \p_a \phi .
\eea
The variation of $m$ in the dust action  gives the  equation of motion
\be
\label{Meom}
m = \pm \frac{P_\phi}{\sqrt{q(q^{ab}\p_a \phi \p_b \phi +1 )}} .
\ee
Using this we eliminate $m$ from the Hamiltonian by writing the dust part of the scalar constraint as
\be
\label{sign}
\mathcal{H}^D = \pm P_\phi \sqrt{q^{ab}\p_a \phi \p_b \phi +1} .
\ee
The sign ambiguity will be determined below when we fix a time gauge.  

\subsection{Imposing circular symmetry}

Let us now impose circular symmetry. A parametrization of the ADM phase space variables $(q_{ab}, \tilde{\pi}^{ab})$ for circular symmetry may be prescribed by using the flat 2D metric   
$e_{ab} dx^a dx^b = dr^2 + r^2 d\theta^2$ and the radial vector $s^a = [1,0] $ and $s_a =s^be_{ab} = [1,0]$. In these coordinates 
\bea
q_{ab}(t,r) &=& \Omega^2(t,r)) s_a s_b + \frac{\rho^2(t,r)}{r^2} (e_{ab}(r) - s_a s_b)   \\
\tilde{\pi}^{ab}(t,r) &=& \frac{P_\Omega(t,r)}{2\Omega(t,r)} s^a s^b + \frac{r^2P_\rho(t,r)}{2\rho(t,r)}\left(e^{ab}(r) -s^a s^b \right).  
\eea
With these definitions the symplectic term in the gravitational action is  
\be
\frac{1}{2\pi} \int dr \ d\theta\ dt \ \tilde{\pi}^{ab} \dot{q}_{ab}  = \int dr \ dt \  \left( P_\rho \dot{\rho} + P_\Omega \dot{\Omega} \right).
\ee
 The ADM metric becomes 
\be
ds^2 = -\left( N^2 - (\Omega N^r)^2 \right) dt^2 + 2 \Omega^2 N^r dr \ dt + \Omega^2 dr^2 + \rho^2 d \theta^2 ,
\ee
and  the Ricci scalar on the slice is 
\be
\label{Ricciscalar}
^{(2)}R = -\frac{2}{\Omega \rho}\left( \frac{\rho'}{\Omega}\right)^\prime.
\ee
Adding the gravitational and dust parts, with  $\sqrt{q} = |\Omega \rho|$, gives the  symmetry-reduced action  
\bea
\label{circular_action}
S &=& \int dr \ dt \  \left( P_\rho \dot{\rho} + P_\Omega \dot{\Omega} + P_\phi \dot{\phi} - N \mathcal{H} - N^r \mathcal{C}_r \right) ,\\
\mathcal{H} &=&  \sgn(\Omega \rho) \left( 2\left(\frac{\rho'}{\Omega}\right)' -\frac{1}{2}P_\Omega P_\rho \right) + 2 \Lambda |\Omega \rho| \pm P_\phi \sqrt{ \left(\frac{\phi'}{\Omega} \right)^2 + 1} ,\\
\mathcal{C}_r &=&  \rho' P_\rho - \Omega P_\Omega^\prime + P_\phi \phi' , 
\eea
where we have used `primes' to denote derivatives with respect to the radial coordinate. As one would expect,  the angular component of the diffeomorphism constraint is identically zero ($\mathcal{C}_\phi \equiv 0$) so that only radial diffeomorphisms play a role in the symmetry-reduced theory. 

The Poisson algebra of the constraints is first class
\bea
\left\{ H(N), H(M) \right\} &=& C_r(NM' - MN') ,\\
\left\{ C_r(N^r), C_r(M^r) \right\} &=& C_r\left(N^r (M^r)' - M^r (N^r)'\right) , \\
\left\{ H(N), C_r(N^r) \right\} &=& - H(N' N^r) ,
\eea
being the reduced version of the Dirac/ADM algebra. At this point, with gauge freedom remaining, there are three pairs of conjugate variables parameterizing the six-dimensional phase space. The physical theory obtained by a Dirac gauge reduction, which fixes the constraints and removes the gauge ambiguity, will leave only one pair of conjugate variables in the two-dimensional physical phase space. In the following we consider the case of non-compact spatial slices with full gauge fixing and appropriate boundary terms to obtain a well-defined variational principle for the canonical 2+1 action. 

\section{Gauge fixing and physical Hamiltonian}

In this section our goal is to obtain the Hamiltonian theory of the  local physical degrees of freedom by  fixing  gauges and solving the constraints.  

We first fix  the radial coordinate gauge by imposing $\chi_\rho:=\rho-r \approx 0$. This is a standard choice in spherical symmetry, and is second class with the diffeomorphsim constraint
\be
\left\{ \chi_\rho, C_r(N^r) \right\} = N^r.
\ee
Keeping this constraint preserved dynamically gives a relation between the lapse and shift functions, 
\be
\label{C1}
N^r - \sgn(\Omega) \frac{N P_\Omega }{2} = 0.
\ee
Solving the  diffeomorphism constraint and imposing this gauge condition removes $\rho$ and $P_\rho$ from the system. We have   
\be
P_\rho = \Omega P_\Omega^\prime -  P_\phi \phi^\prime,
\ee
which leads to the the partially gauge-fixed action
\bea
S &=& \int dr \ dt \  \left( P_\Omega \dot{\Omega} + P_\phi \dot{\phi} - N \mathcal{H} \right) ,\\
\mathcal{H} &=& \frac{1}{2} \sgn(\Omega) P_\Omega \left(  P_\phi \phi^\prime -\Omega P_\Omega^\prime \right) + 2 \Lambda |\Omega| r + \left(\frac{2}{|\Omega|}\right)'  \pm P_\phi \sqrt{ \left(\frac{\phi'}{\Omega} \right)^2 + 1} .
\eea
In writing the Hamiltonian we assume that  $\sgn(\Omega)$  is fixed since we must have $|\Omega(t,r)|>0$ for the metric be non-degenerate.

We now choose the dust time gauge by adding the constraint $\chi_\phi:= \phi - t \approx 0$, a condition which is  second class with the Hamiltonian constraint:
\be
\left\{ \chi_\phi, H(N) \right\} = \pm N .
\ee
Requiring that this gauge is dynamically preserved leads to 
\be
\label{C2}
N=\pm 1.
\ee
Recalling now the definition of the momentum $P_\phi$ (\ref{P-phi}), we see that the signs of $m$ and $N$ are linked in this gauge by $m \sqrt{q}= N P_\phi$. Therefore choosing $N=1$, which generates dynamics forward in time, fixes \cite{Swiezewski:2013} the sign ambiguity arising from (\ref{Meom}): $P_\phi = +m\sqrt{q}$ and the solution of the Hamiltonian constraint gives  the physical Hamiltonian density 
\be
\label{Pphi}
-{\cal H}_{\mbox{\scriptsize phys}}= P_\phi = 
 |\Omega| \left(\frac{P^2_\Omega}{4}\right)^\prime - 2\Lambda |\Omega| r - \left(\frac{2}{|\Omega|}\right)'.
\ee
The shift function is also fixed  via (\ref{C1}): 
\be
\label{lapse+shift}
N=+1 \iff N^r = \sgn(\Omega) \frac{P_\Omega}{2} .
\ee

\subsection{Reduced action}

The reduced action is obtained by substituting the gauge fixing conditions and the solutions of the constraint (\ref{Pphi}) into the starting action (\ref{circular_action}). This gives
\be
\label{red-S}
 S= \int dt \int_\Sigma dr \  \left( P_\Omega \dot{\Omega}  - {\cal H}_{\mbox{\scriptsize phys}} \right) - \int_{\p \Sigma} dt \  \frac{2}{|\Omega|}, 
\ee
where 
\be
\label{Hphys}
\mathcal{H}_{\mbox{\scriptsize phys}} = 2 \Lambda |\Omega| r - |\Omega| \left(\frac{P^2_\Omega}{4}\right)^\prime.
\ee
The boundary term  arises from the total derivative present in (\ref{Pphi}), and comes ultimately from the Ricci scalar density $\sqrt{q}^{(2)}R = (2/|\Omega|)'$.  

If  $\Sigma_t$ is asymptotically flat, as will be the case for some but not all solutions to the equations of motion, this term evaluated at a fixed radius
\be
 \left. \frac{2}{|\Omega|} \right|_{r=r_0}
\ee
determines the energy within a disc of radius $r_0$ \cite{Ashtekar:1994}, and as we shall see below, this term also gives a measure of deficit angle at the origin in the limit $r_0 \rightarrow 0$. This is because in 3D gravity a conical defect represents a point source of energy at the origin;  the relationship between deficit angle $\alpha$ and the energy $M$ of the point source (in units $8G = 1)$ is $M=\alpha/\pi$ \cite{Deser:1983tn}.

Many interesting solutions in 3D gravity are singular at the origin \cite{Carlip:1994ap}, and in order to allow for these solutions we excise the origin $r=0$. This ensures that the metric and curvature are well-defined everywhere on each $\Sigma_t$. We handle this excision by restricting the radial coordinate to the range $r\in (0, r_{\mbox{\scriptsize max}}]$.  Thus each spatial slice $\Sigma_t$ is taken to have an outer and an inner boundary.

With these considerations in hand, we turn to a discussion of the  functional differentiability of the action. This requires specifying what  variations are to be fixed on the boundaries, and may require the addition of more boundary terms \cite{Regge:1974zd}.   Variation of the action (\ref{red-S}) with respect to $\Omega$  gives the boundary terms 
\be
\lim_{\epsilon \rightarrow 0} \int_\epsilon^{r_{\mbox{\tiny max}} } dr \Big[ P_\Omega \delta \Omega \Big]^{t=t_f}_{t=t_i}
\ee 
for some initial and final times $t_i$ and $t_f$, and 
\be
\lim_{\epsilon \rightarrow 0} \int_{t_i}^{t_f} dt \ \left[ \frac{2}{\Omega^2} \delta |\Omega| \right]^{r=r_{\mbox{\tiny max}}}_{r=\epsilon}.
\ee
And variation with respect to $P_\Omega$ gives the boundary term
\be
\lim_{\epsilon \rightarrow 0}  \int dt\left[  \frac{|\Omega|}{2} P_\Omega \delta P_\Omega \right]_{r=\epsilon}^{r=r_{\mbox{\tiny max}}}. 
\ee

We define the variational principle by fixing $\Omega$ at the end points by 
\be
 \Omega(t, r_{\mbox{\tiny max}}) = a(t), \ \ \ \ \ \ \ \ \  \Omega(t, \epsilon) = b(t). 
\ee
With this choice the $\delta \Omega$ variation is well defined.  The surface term arising from the symplectic piece is zero  because initial data and its subsequent evolution fix $\Omega$ at $t_i$ and $t_f$.  Lastly,  to keep $P_\Omega$  free at the boundaries, we add a surface term to cancel the $\delta P_\Omega$ variation.  The final gauge fixed action is 
\be
\label{red-S-final}
 S= \int dt \int_\Sigma dr \  \left( P_\Omega \dot{\Omega}  - {\cal H}_{\mbox{\scriptsize phys}} \right)    - \lim_{\epsilon \rightarrow 0}  \int dt \left [\frac{2}{|\Omega|} +  \frac{1}{4}|\Omega| P_\Omega^2  \right]_{r=\epsilon}^{r=r_{\mbox{\tiny max}}}. 
\ee

The  summary so far is that in the process of deriving this action,  the dust field and its conjugate momentum have been eliminated from the system, and the remaining metric field and its conjugate momentum $(\Omega, P_\Omega)$  describe the  geometry. Furthermore, as noted in \cite{Husain:2011tk}, the dust time gauge results in the conversion of the former hamiltonian constraint of pure gravity into a non-vanishing true hamiltonian.  

There are other possibilities for fixing the variational principle. For example, we could have gone without adding the second boundary term and instead fixed the momentum $P_\Omega$ on the boundaries. The choice above is the simplest since it requires boundary conditions for $\Omega$ only, and still permits a large class of interesting solutions.  
 
With the variational principle well-defined, the equations of motions are 
\bea
\dot{\Omega} &=&  \frac{P_\Omega}{2}  |\Omega|' , \\
\dot{P_\Omega} &=&  \sgn(\Omega) \left(\frac{P_\Omega}{2} P_\Omega^\prime  -  2\Lambda r\right).
\eea

\subsection{Physical conditions}

Let us consider the spacetime metric and physical properties resulting from these gauge choices. In fixing a gauge for the field variables, we obtained conditions  which fix the lapse and shift functions (\ref{lapse+shift}). The resulting metric is 
\be
\label{metric}
ds^2 = -\left(1-\frac{\Omega^2 P_\Omega^2}{4} \right) dt^2 + \sgn(\Omega) \Omega^2 P_\Omega\  drdt + \Omega^2 dr^2 + r^2 d\theta^2.
\ee
The metric is non-degenerate so long as $\Omega^2>0$, which implies that $\sgn(\Omega)$ is constant.
  
\subsubsection{Deficit angle} 

Consider the ratio $\mathcal{F}$ of  the circumference of a circle with radius $r_0>0$ divided by the proper radius
\be
\mathcal{F}:=\lim_{\epsilon \rightarrow 0} \frac{\displaystyle \int_{r=r_0+\epsilon} \sqrt{g_{\theta \theta}} \ d\theta }{\displaystyle \int_{\epsilon}^{r_0+\epsilon} \sqrt{g_{rr}}\  dr}.
\ee
In flat space this is $2\pi$, but in general we may have $\mathcal{F} = 2\pi - \alpha(r_0)$ where $\alpha(r_0)$ 
is a deficit angle given by 
\be
\alpha(r_0) = 2\pi\left(1 -  \frac{r_0}{\displaystyle \lim_{\epsilon \rightarrow 0} \int_{\epsilon}^{r_0+\epsilon} |\Omega| dr} \right). \label{def-angle}
\ee

The limit of vanishing radius $r_0$, after first taking the limit $\epsilon \rightarrow 0$, gives  $\alpha = 2\pi\left(1 - \frac{1}{|\Omega(t,0)|} \right)$. This implies that constant-time slices of the above metric generically describe a conical geometry near the origin with a time-dependent deficit angle. Note that in 3D gravity, a negative deficit angle corresponds to a point source with negative energy \cite{Deser:1983tn, Carlip:1998uc}. In order to have a positive semi-definite energy at the origin, one would require that
\be
|\Omega(t,0)|\ge 1 .
\ee

\subsubsection{Energy density} 

The stress-energy tensor may also be written in terms of the phase space variables.  From the action we find that
\be
T_{\mu \nu} := - \frac{2}{\sqrt{-g}} \frac{\delta S_D}{\delta g^{\mu \nu}} = \frac{m}{2 \pi} \delta^t_\mu \delta^t_\nu.
\ee
We see that $m$ is the time-time component of the above. From the metric, or using (\ref{Pphi}) and (\ref{Meom}) with the positive choice of sign to solve for $m$, we obtain the  energy density 
\be
\label{energy}
2 \pi \mathfrak{T}_{tt} = \sqrt{|g|} m
= |\Omega| \dot{P}_\Omega - \left(\frac{2}{|\Omega|}\right)^\prime,
\ee
where the equations of motion (\ref{Oeom}, \ref{Peom}) have been used to simplify the expression. A positive definite energy density requires that the right hand side be greater than or equal to zero.  The spacetime Ricci scalar is: 
\be
\label{Ricci3}
^{(3)}R = m = \frac{\dot{P}_\Omega}{r} - \frac{1}{r} \left(\frac{1}{\Omega^2}\right)^\prime .
\ee
Since we have excised the point $r=0$ from the spatial manifold, the curvature scalar is missing a delta function contribution at the origin when there is a conical singularity.

\subsubsection{ Horizons}  

Congruences of future directed outgoing and ingoing radial null geodesics are  
%
\be
u^\mu = \left( 1, \sgn(\Omega)\left(\frac{1}{\Omega} - \frac{P_\Omega}{2}\right), 0 \right) , \qquad \qquad  v^\mu = \left( 1, -\sgn(\Omega)\left(\frac{1}{\Omega} - \frac{P_\Omega}{2}\right), 0 \right) .
\ee
These satisfy $u^\mu v_\mu = -2$, and provide the null expansions. In our context these are  {\it physical} phase space observables which are potentially useful in a quantum theory \cite{Husain:2004yy}.  
The outward null expansion of circles embedded in a spatial slice with unit  outward normal  $s^\mu = (0, 1, 0)/\Omega$ is 
\be
\label{expansion}
\Theta := \left(q^{\mu \nu} -   s^\mu s^\nu \right) \nabla_\mu u_\nu = \frac{1}{r}\left(\frac{P_\Omega}{2}-\frac{1}{|\Omega|} \right).
\ee
Dynamical apparent horizons are obtained by solving $\Theta(t,r)=0$ to give the horizon radius $r_h(t)$. This may have multiple solutions on a given time slice (see eg. \cite{Husain:1994uj, Ziprick:2009} for explicit examples).

\section{Classical solutions}

In this section we discuss classical solutions to our model. We find a large class of exact solutions and provide several examples. In particular we obtain a static solution for $\Lambda < 0$ which describes the BTZ black hole, and for $\Lambda = 0$ we find travelling wave solutions.

To this point we have left the sign of $\Omega$ arbitrary. As noted above, we must require that $\sgn(\Omega)$ is constant throughout the spacetime in order for the action to be well-defined. This implies that the solution space is split into sectors with $\sgn(\Omega) = \pm 1$. To keep the presentation simple, we assume $\Omega > 0$ for the remainder of the article. The solution space for $\Omega < 0$ is nearly identical with only trivial differences.

\subsection{$\Lambda\ne 0$}

For the case of non-zero cosmological constant the equations of motion are:
\bea
\label{Oeom}
\dot{\Omega} &=& \frac{P_\Omega}{2}  \Omega'  , \\
\label{Peom}
\dot{P_\Omega} &=& \frac{P_\Omega}{2} P_\Omega^\prime  -  2\Lambda r .
\eea
The second equation is similar to inviscid Burger's equation, but with a source term coming from the cosmological constant; it contains only the momentum and can be solved independently. This is coupled to the first equation which resembles the advection equation but with a variable speed of propagation given by $P_\Omega/2$. As we will see, any solution for the momentum then determines how initial data $\Omega(0,r)$ evolves.

\subsubsection{General solution}

There is an auxilliary, flat spacetime with Lorentzian signature defined by the $(t,r)$ plane. On the auxilliary spacetime the momentum equation (\ref{Peom}) is a conservation equation $\p_a j_1^a = 0$ for the current:
\be
j_1^a = \left[-P_\Omega,\ \frac{P_\Omega^2}{4} - \Lambda r^2 \right],
\ee
which has an associated conserved charge given by:
\be
q_1 = \lim_{\epsilon \rightarrow 0} \int_\epsilon^{r_{\mbox{\tiny max}}}  P_\Omega \ dr .
\ee
Considering the system as a whole there is another conserved current:
\be
j^a_2 = \left[ \Omega \left(2 \Lambda r - \frac{P_\Omega}{2} P_\Omega^\prime \right) , \  \Omega \frac{P_\Omega}{2}  \left( \frac{P_\Omega}{2} P_\Omega^\prime - 2 \Lambda r \right)\right],
\ee
where the conserved charge is:
\be
q_2 = \lim_{\epsilon \rightarrow 0} \int_\epsilon^{r_{\mbox{\tiny max}}} \Omega \left(\frac{P_\Omega}{2} P_\Omega^\prime - 2 \Lambda r  \right) \ dr .
\ee

It is well-known that conservative equations can be solved by the method of characteristics. To employ this method, we consider characteristic lines parameterized by $s$, described in terms of parametric equations for the coordinates $r(s)$ and $t(s)$. Differentiating with respect to $s$, we obtain:
\be
\frac{d}{ds} P_\Omega = \dot{P}_\Omega \frac{\p t}{\p s} + P_\Omega^\prime \frac{\p r}{\p s} = -2 \Lambda r .
\ee
This is equivalent to the equation of motion if we have the following equations along each characteristic:
\be
\frac{\p t}{\p s} = 1, \qquad \qquad \frac{\p r}{\p s} = - \frac{P_\Omega}{2}, \qquad \qquad \frac{d}{ds} P_\Omega = -2 \Lambda r.
\ee
These equations are solved by:
\be
\label{parametric}
t = s, \qquad \qquad r = r_0 \cosh \sqrt{\Lambda}s - \frac{P_0}{2 \sqrt{\Lambda}} \sinh \sqrt{\Lambda} s, \qquad \qquad P_\Omega =  P_0\cosh \sqrt{\Lambda}s - 2\sqrt{\Lambda} r_0 \sinh \sqrt{\Lambda} s ,
\ee
where the initial values are $r_0 = r(s=0)$ and $P_0 = P_\Omega(s=0)$. Each characteristic is labelled by the `starting point' $r_0$, and initial data for the momentum is a function of the radial points on the initial slice $P_\Omega(t=0, r) = P_0(r_0)$.


Given a solution for the momentum we can solve (\ref{Oeom}) for $\Omega$ using the characteristic method again. The characteristics for this equation are the same as those for the momentum equation of motion, but here we have:
\be
\frac{d}{ds} \Omega = 0,
\ee
so that $\Omega$ is constant along each characteristic. This implies that given some initial data $\Omega(0,r)$, the configuration field simply flows along the characterstic lines defined by the momentum.

\subsubsection{Examples}

Static solutions are obtained  by setting $\dot{\Omega} = \dot{P_\Omega} = 0$ in (\ref{Oeom}, \ref{Peom}). This gives 
\bea
\Omega &=& C_1 , \\
P_\Omega &=& \pm  2\sqrt{C_2 + \Lambda r^2}.
\eea

The metric may be put in  a more succinct form as follows. Rescale $C_2$ and the $r$ and $\theta$ coordinates by absorbing the constant $C_1$  as $\t{r} = C_1 r$, $\t{\theta} = \theta/C_1$ and $\t{C}_2 = C_1^2 C_2$. With this rescaling the angular coordinate has a range $0\le \t{\theta} \le 2\pi / C_1$ so that a choice of $C_1>0$ implies a deficit angle defined by $\Omega$ as described in the preceeding section. The line element becomes 
\be
ds^2 = - f dt^2 \pm 2 \sqrt{1-f} \  d\t{r} \ dt  + d\t{r}^2 + \t{r}^2 d\t{\theta}^2, 
\ee
where $f(\t{r}) \equiv 1 - \t{C}_2 - \Lambda\t{r}^2$.

This solution remains well-defined for any choice of $C_2 > 0$, which is required for $P_\Omega$ to be real at each point. For de Sitter spacetime ($\Lambda>0$) there are no additional restrictions, but for the anti-de Sitter (AdS) case ($\Lambda<0$) the radial coordinate has a limited extent in order to keep $P_\Omega$ non-imaginary:
\be
0 < \t{r} \le \sqrt{\frac{\t{C}_2}{|\Lambda|}}.
\ee
Let us consider the AdS case further. The above line element is in fact a generalization of the BTZ black hole which allows for a deficit angle due to the choice of $C_1$. This can be seen by transforming to a new time coordinate 
\be
\t{t} = t \pm \int_{0}^{\t{r}} \frac{\sqrt{1-f(x)}}{f(x)} dx,
\ee
which puts the line element in the form 
\be
ds^2 = -f d\t{t}^2 + f^{-1} d\t{r}^2 + \t{r}^2 d\t{\theta}^2 , 
\ee
where we note again that the angular range is $0 \le \t{\theta} < 2\pi/C_1$. This spacetime has an event horizon at $\t{r} = \sqrt{(\t{C_2} - 1)/|\Lambda|}$, and when $C_1 = 1$ it is the BTZ spacetime in flat slice coordinates.
 
\subsection{$\Lambda = 0$}

With zero cosmological constant the equations of motion have a remarkable symmetric form
\bea
\label{eom1}
\dot{\Omega} &=&  \frac{1}{2} P_\Omega \Omega' ,  \\
\label{eom2}
\dot{P_\Omega} &=& \frac{1}{2}P_\Omega P_\Omega' .    
\eea
The momentum equation of motion is now Burger's equation with vanishing viscosity. There is a substantial volume of literature on the subject. Most notably, this equation gives shock waves when characteristics cross.

\subsubsection{General solution}

The equations of motion are again conservation equations $\p_a j^a=0$ in the auxiliary flat  Lorentzian spacetime defined by the $(t,r)$ plane. With $\Lambda = 0$ the currents are
\be
j_1^a = \left[ -P_\Omega, \ \frac{P_\Omega^2}{4} \right], \qquad \qquad j^a_2 = \left[-\Omega \left(\frac{P_\Omega^2}{4}\right)^\prime, \ \Omega \frac{P_\Omega^2}{4} P_\Omega^\prime \right],
\ee 
and the corresponding conserved charges are
\be
q_1 = \lim_{\epsilon \rightarrow 0} \int_\epsilon^{r_{\mbox{\tiny max}}}  P_\Omega \ dr  , \qquad \qquad q_2 =  \lim_{\epsilon \rightarrow 0} \frac{1}{2} \int_\epsilon^{r_{\mbox{\tiny max}}} \Omega P_\Omega P_\Omega^\prime \ dr .
\ee 

The equation of motions can again be solved by the methods of characteristics; one need only put $\Lambda=0$ in the equations from the last section. There are two classes of solutions: 1) $P_\Omega$ is constant and $\Omega = h\left(r + \frac{P_\Omega}{2}t \right)$; 2) $P_\Omega$ is not constant and $\Omega = h\left(P_\Omega\right)$, for an arbitrary function $h$. This last fact is a remarkable consequence of the structure of the $\Lambda=0$ equations. When $P_\Omega$ is constant the characteristics are guaranteed not to cross.

We also note that for vanishing cosmological constant the parametric equations (\ref{parametric}) can be inverted to yield the following:
\be
\label{roots}
2r + P_\Omega t - 2 f(P_\Omega) = 0.
\ee
Given any choice of function $f(P_\Omega)$ of the momentum, solutions to $(\ref{eom2})$ are given by the roots to this equation.

\subsubsection{Examples}

Let us note three types of  solutions. The first is a class of static solutions obtained by setting $P_\Omega=0$ in (\ref{eom1}-\ref{eom2}). This implies $\Omega= f(r)$, a nowhere-vanishing, but otherwise arbitrary function. The resulting metric is 
\be
ds^2 = -dt^2 + f(r)^2 dr^2 + r^2 d\theta^2.  
\ee
The energy density is given by $\displaystyle 2 \pi \mathfrak{T}_{tt} = \frac{2 f'}{f^2}$, and its sign is determined by the sign of $f'$. The spacetime Ricci scalar is $\displaystyle ^{(3)}R=\frac{f'}{r f^3}$, and the $r^{-1}$ factor indicates that solutions are generally singular at $r=0$, except for the particular choice $f(r) = \pm (C_1 - C_2 r^2)^{-1/2}$. Constant time slices are cones with deficit angle $\alpha = 2 \pi (1 - 1/f(0))$, and there are no horizons in this spacetime.

The second is a self-similar solution obtained by setting $f(P_\Omega)=0$ in (\ref{roots}) to obtain
\be
P_\Omega = - \frac{2r}{t}. 
\ee
It is immediate that this solves (\ref{eom2}), and leads to the $\Omega$ equation of motion 
\be
\dot{\Omega} = - \frac{r}{t}  \Omega'. 
\ee
One solution of this is $\Omega=1$, which gives the metric 
\be
ds^2 = -\left(1- \frac{r^2}{t^2}  \right)dt^2 - \frac{2r}{t} drdt + dr^2 + r^2 d\theta^2. 
\ee
 The Ricci scalar from (\ref{Ricci3}) is $^{(3)}R = 2/t^2$ so there is a spacelike curvature singularity  at $t=0$. Looking at the condition (\ref{expansion}) we find horizons at $r=- t$. Constant $t$ slices are flat without any deficit angle due to the choice of $\Omega=1$. 
  
A third class of solutions is obtained by setting $P_\Omega = 2v$ for some constant $v \in \mathbb{R}$. The $\Omega$ equation reduces to the   advection equation
\be
\dot{\Omega} = v \Omega'\ ,
\ee
which has the general solution
\be
\Omega = h(r + v t) \equiv h(u)
\ee
for arbitrary $h$ and no restriction on $v$, where $u \equiv r+vt$ labels each (straight) characteristic. If we choose $\Omega = C$, a constant, we have a flat metric with deficit angle $\alpha = 2\pi(1- 1/C)$ as described in (\ref{def-angle}).

For a non-constant $\Omega$, the $v>0$ and $v<0$ solutions describe respectively radially ingoing and outgoing profiles.  The $v>0$ wave metric is 
\be
ds^2 = -\left[1-(v\Omega)^2 \right] dt^2+  2v\Omega^2   drdt + \Omega^2 dr^2 + r^2 d\theta^2. \label{wave-met}
\ee
We note the following features of these ``wave" solutions. The Ricci scalar from (\ref{Ricci3}) is 
\be
\ ^{(3)}R = \frac{2}{r} \left( \ln \Omega\right)' = \frac{2}{rh} \frac{dh}{du},
\ee
and there are dynamical horizons if $\Omega(t,r) = 1/v$. Thus horizons will be present if the maximum value of the wave profile exceeds $1/v$, and the minimum is less than $1/v$.

From the expression for energy density (\ref{energy}) we see that the energy flux is positive where $\Omega'>0$ and negative where $\Omega'<0$. There is a conical singularity at the origin with deficit angle $\alpha = 2 \pi (1 - 1 / \Omega(t,0))$. As the energy flux reaches the origin, positive flux adds to the deficit angle (which represents the mass of the singularity), and negative flux reduces the deficit angle.

\section{Quantum theory}

In this section we describe a non-perturbative quantization of the $\Lambda = 0$ theory. The circularly symmetric sector of the model we are considering has one local physical degree of freedom $(\Omega)$,  and as we have shown in the last section, the classical theory can be solved by the method of characteristics with the $P_\Omega$ solution providing a local $(t,r)$-dependent speed for the $\Omega$ equation.  The full quantum theory of this sector is more challenging. Although there is a physical hamiltonian and no constraints, the hamiltonian is unconventional in the sense that there is no separation of pure kinetic and potential terms. We can however achieve a full quantization of the $P_\Omega = \mbox{constant}$ sector of the solution space discussed in the last section. 

As we noted, this sector of the solution space describes either purely ingoing or outgoing waves. A first challenge is that since $P_\Omega$ is a constant, the symplectic structure we have been using up to now is not available. This is overcome by noting that we can obtain a new symplectic structure starting from the solution space of a differential equation \cite{Crnkovic:1986ex}. The basic idea involves defining geometric structures on the solutions space that leads to a conserved symplectic current. The integral of this current  on an initial value surface defines the desired symplectic form.

For completeness, we now derive the symplectic two-form for the circularly symmetric, $\Lambda = 0$, $P_\Omega = 2v$ (where $v$ is a constant), sector of the solution space and refer the reader to \cite{Crnkovic:1986ex} for details. Once the symplectic structure is obtained, we move on to the canoncial quantum theory.

In our case the differential equation is
\be
\label{eq}
\dot{\Omega} = v \Omega' ,
\ee
on the half plane $t\in (-\infty,\infty)$, $r\in (0,\infty)$. For our purposes, we take this half-plane to define an auxilliary spacetime $M$ with a flat Lorenzian metric given by $\eta = \mbox{diag}(-v^2,1)$.

Consider the space $Z$ of solutions to (\ref{eq}). A point $\Omega \in Z$ represents a solution to this equation, and a tangent vector $\delta \Omega$ at this point is a small displacement which must also be in the solution space $Z$. Writing the displacement of the solution as $\Omega + \delta \Omega$, we find that the tangent vector $\delta \Omega$ must also satisfy (\ref{eq}).

The space of one-forms on $Z$ is dual to the tangent space. If we label the spacetime points $x \in M$, then the one-form dual to $\delta \Omega$ is given by $\delta \Omega(x)$. It is important to note that these one-forms are anti-commutative, 
\be
\delta \Omega(x) \ \delta \Omega(y) + \delta \Omega(y) \ \delta \Omega(x) = 0.
\ee
The symplectic current is defined by 
\be
J_a(x) = \delta \Omega (x) \p_a \delta \Omega(x),
\ee
which due to the equation of motion (\ref{eq}) and the anti-commutivity of one-forms, obeys the conservation equation $\eta^{ab}\p_a J_b = 0$. The associated conserved charge is given by integrating over a spatial hypersurface:
\be
\omega = \int dr \ J_t = \int dr \ \delta \Omega \ \delta \dot{\Omega} . 
\ee
This conserved charge is the symplectic two-form we seek. It implies that the momentum conjugate to $\Omega$ is  $\Pi:= \dot \Omega$, with the equal-time Poisson algebra
\be
\label{PBs}
\left\{ \Omega(r,t), \Pi(r', t)   \right\} = \delta(r - r'), \quad
\left\{  \Omega(r,t), \Omega(r', t)   \right\} = \left\{\Pi(r,t), \Pi(r', t )  \right\} = 0.
\ee
These are equivalent to the Poisson brackets for free scalar field theory, but the Hilbert space we define next will differ in that it includes only the ingoing \textit{or} the outgoing modes, depending on the sign of $v$ chosen.

Having obtained the symplectic structure, let us consider the Hilbert space we will use for quantization.  Consider the positive energy (dust time) mode functions
\be
\psi^\pm_k(r,t) = e^{-ivkt} \left(e^{-ikr} \pm e^{ikr} \right), \ \ \ k > 0.
\ee
These sets have different boundary conditions at $r=0$:  $\psi^+(t,r=0)= 2e^{-ivkt}$  and $\psi^-(t,r=0)= 0$.  Both sets are orthogonal  and complete on the half-line:
\bea
\int_0^\infty  dr\  \bar{\psi}^\pm_k (r,t) \ \psi^\pm_{k'}(r,t)  &=& 2\pi \delta(k-k'),\nn\\
\int_0^\infty dk\    \bar{\psi}^\pm_k (r,t) \  \psi^\pm_k(r',t)  &=&  2\pi \delta(r-r'),
\eea
and also satisfy 
\be
 \int_0^\infty dr\  \bar{\psi}^\pm_k (r,t) \ \psi^\mp_{k'}(r,t) =0.  
\ee

Let ${\cal H}^\pm_v$ denote the Hilbert spaces with the bases $\psi^{\pm}_k$, and let ${\cal H}_v= {\cal H}^+_v \oplus {\cal H}^-_v$.  The purely ingoing (outgoing) wave solutions  are obtained by the normalized linear combination
\be
  g_k(r,t) :=  \frac{1}{\sqrt{2\pi}} \left(  \psi^+_k(r,t) + \psi^-_k(r,t) \right) =  \frac{1}{\sqrt{\pi}} e^{-ik(vt+r)}. \label{inmodes}
\ee
Clearly $g_k \in {\cal H}_v$ are solutions of our model. They may be viewed as ``quasi-particles" if the bases given above for  ${\cal H}^+$ and ${\cal H}^-$ are viewed as ``particles."  The Hilbert space of the entire wave sector (all ingoing and outgoing modes labelled by $v\in \mathbb{R}$)  is the tensor product 
\be
{\cal H} = \otimes_v {\cal H}_v \label{fullH}
\ee

Let us  demonstrate the quantization in the  $v=1$ component.   $\Omega(r,t)$ and its conjugate $\Pi(r,t)$ may be represented as operators in ${\cal H}_1$, in the manner that is standard in field theory:
\bea
\hat{\Omega} (r,t) &=& \int_0^\infty dk \ \frac{1}{\sqrt{k}}\left( \hat{a}_k g_k(r,t) + \hat{a}_k^\dagger \bar{g}_k (r,t) \right), \\
\hat{\Pi} (r,t) &=& - i \int_0^\infty dk \ \sqrt{k}\left( \hat{a}_k g_k(r,t) - \hat{a}_k^\dagger \bar{g}_k (r,t) \right).
\eea
Their commutator algebra implied by the Poisson algebra (\ref{PBs}) leads to the  usual commutators for ladder operators
\be
[\hat{a}^\dagger_k, \hat{a}^\dagger_{k'}] = [\hat{a}_k, \hat{a}_{k'}] = 0 \qquad \qquad [\hat{a}_k, \hat{a}^\dagger_{k'}] =  \delta(k-k').
\ee
  Fock states  are constructed by starting with the vacuum state $|0_k\rangle$ defined by  $\hat{a}_k|0_k\rangle = 0$, and the $n_k-$particle states by 
\be
|n_k\rangle = \frac{1}{\sqrt{n_k}}\left[\hat{a}_k^\dagger\right]^{n_k}|0_k\rangle. 
\ee
The Fock basis is given by products of the $n_k$-particle states with different $k$ values.
 
This completes the specification of the quantum theory  for the $P_\Omega=$ constant sector of the solution space.  What it shows is that this  non-perturbative sector of 2+1 gravity coupled to pressureless dust  in spherical symmetry is  {\it exactly dual} to the quantum theory of a massless scalar field on the half line.

\subsection{Metric operator}   
     
With the above quantization  we can now proceed to describe the ``quantum geometries"  for this spherically symmetric sector of the model. The metric contains only the function $\Omega$ so it is possible to define the metric operator in the Hilbert space ${\cal H}$ by $ \hat{g}_{ab} := g_{ab}(\hat{\Omega})$. A notion of geometry is given by the expectation value of this operator in a quantum state. There are thus an infinite set of possible geometries depending on the choice of state.

The metric contains the factor $\hat{\Omega}^2$ so we need to select an operator ordering of $\hat{a}_k,\hat{a}_k^\dagger$ to define it. This is provided by  imposing the physical requirement that expectation values in semiclassical states give recognizable classical solutions. One choice for such states are the coherent states defined by 
\be
\hat{a}_k|\alpha_k\rangle = \alpha_k|\alpha_k\rangle
\ee
These states are explicitly given by  \cite{Glauber:1963tx}
\be
|\alpha_k\rangle = e^{-|\alpha_k|^2 /2} \sum_{n_k=0}^\infty \frac{\alpha_k^{n_k}}{\sqrt{n_k!}} \ |n_k\rangle.
\ee
It is known that the expectation values of normal ordered operator in these states give the corresponding classical results. We therefore define
\be
\hat{g}_{ab} := g_{ab}(: \widehat{\Omega^2} :) \ .
\ee
As an explicit example let us consider the state which is the vacuum for all modes except $k$, and the coherent state for mode $k$,  
\be
|\psi\rangle = |\alpha_k\rangle \prod_{j\ne k} |0_j\rangle.
\ee
This gives
\be
\langle :\widehat{\Omega^2}: \rangle = \frac{1}{k} \left((g_k(r,t))^2\  \alpha_k^2 + (g_k^*(r,t))^2  (\alpha_k^*)^2 + 2 |\alpha_k|^2 \right),
\ee
where $\alpha_k$ are any complex numbers specifying a classical solution. The complex number $\alpha_k$ must be such that the right hand side is positive definite in order to avoid a degenerate metric, and depending upon the value of $\alpha_k$ there may be apparent horizons. The quantum fluctuations in these states 
\be
\Delta (\Omega^2) = \langle \: \widehat{\Omega^2} :\  : \widehat{\Omega^2}:   \rangle -  \langle : \widehat{\Omega^2}:  \rangle^2  
\ee
is of course not zero, since $|\alpha_k\rangle$  are the minimum uncertainty states. 

The expectation value of the metric in the $n_k$-particle state $|n_k\rangle$ on the other hand gives the metric 
\be
ds^2 = - \left(1-\frac{2n_k}{k}\right) dt^2 + \frac{4n_k}{k}drdt + \frac{2n_k}{k}dr^2 + r^2 d\theta^2.
\ee
The constant time slices are cones with deficit angle
\be
\alpha = 2 \pi \left(1 - \sqrt{\frac{k}{2n_k}} \right).
\ee
Recall that in 3D gravity, a conical singularity corresponds to a point source with a mass proportional to the deficit angle \cite{Deser:1983tn, Carlip:1998uc}. This implies a discrete mass spectrum of the $n_k$-particle states determined by the wave number $k$. It asymptotes  to $2\pi$ as $n_k$ gets large, and has a positive or zero energy for $2 n_k \ge k$.

 For special values of the parameters satisfying $2n_k=k$, the apparent horizon function vanishes everywhere $\Theta(r,t)=0$. With these values, the conical defect of the spacetime slicing is such that the outward going null geodesics remain at constant radius.
 
The self-adjoint metric operator defined above using the creation and annihilation operators provides, via the expectation value,  a correspondence between quantum states and spacetime geometries. The  coherent states lead to classical geometries with fluctuations. There is also the interesting possibility of obtaining ``macroscopically entangled geometries'' by using states that describe entangled superpositions.   Construction of such states requires either two systems described  by a tensor product Hilbert space (such as that for two spin one half particles), or  the division of a Hilbert space into two subsystems. 

In the quantum theory of the wave sector we have discussed, it is possible to produce entangled states by considering for example states in the subspace  ${\cal H}_1 \otimes {\cal H}_{-1}$  of the full Hilbert space (\ref{fullH}).  Let  $|\alpha_k\rangle_1$ denote a semiclassical state in ${\cal H}_1$ and  $|\beta_k\rangle_{-1}$ a semi-classical state in ${\cal H}_{-1}$. Then an example of a macroscopic entangled state of spacetime geometries is  
\be
|\psi\rangle  =  \frac{1}{\sqrt{2}} \left(  |\alpha_k\rangle_1     |\beta_k\rangle_{-1} + |\beta_k\rangle_1  |\alpha_k\rangle_{-1}       \right).
\ee
There are numerous examples of this type involving two or more subsystems, even within a fixed $v$ sector of the Hilbert space, but with states labelled by different $k$ values. In a full quantum theory of gravity it would presumably be possible to divide the physical Hilbert space into sectors corresponding to, for example, black hole and cosmological geometries. It would then be possible to construct interpretationally  challenging  macroscopically entangled states.

\subsection{Quantum horizons} 
     
From the forgoing we can consider the idea of  a ``quantum horizon'' defined by a horizon operator \cite{Husain:2004yy}
 \be
   \hat{h} = \ : \widehat{\Omega^2} : - \frac{1}{v^2}I, 
 \ee
 which is the operator analog of the classical apparent horizon condition. In our gauge fixed setting this is physical observable. It is clear that the fluctuation of this operator is non-zero in a coherent state, and so the corresponding dynamical horizons are not sharply defined as they are in the classical theory. For a coherent state with an $\alpha_k$ such that the horizon condition $\langle \hat{h} \rangle = 0$ is marginally satisfied, fluctuations of the metric operator can lead to uncertainty in whether or not horizons are present at all.

\section{Summary and Discussion} 

We studied a new model for gravity in 2+1 dimensions. Unlike most of the existing literature on 3D gravity, the model has a local degree of freedom which manifests itself as a  metric function  in the dust time gauge.  The resulting theory has novel aspects in circular symmetry. The equations of motion are simple yielding several interesting classes of solutions, including waves;  the latter provide a ``midi-superspace" sector of the  solution space  which is amenable to Fock quantization.

The quantization provides some interesting and precise results. Among these is the  observation that horizons fluctuate, which we showed using semiclassical states. It is natural to expect that this result goes over to four dimensions where it could inform issues such as the so called information loss  problem in black hole evaporation. In particular, metric fluctuations imply that the separation of the Hilbert space into states which are strictly  inside / outside the horizon, as is common in computing  entanglement entropy, is an ambiguous procedure. Metric fluctuations further imply that the time of apparent horizon formation may be ambiguous for any choice of time coordinate. 

Metric fluctuations also inform the``firewall" issue, which at its  core requires exactly null non-fluctuating horizons as a fundamental assumption. If a horizon is leaky because it has fluctuations, then it is clear that the central argument based on the impossibility of simultaneous perfect correlation  between modes across a horizon on the one hand, and perfect correlation between early and late time Hawking radiation on the other, ceases to be an issue: no perfect null horizon, no monogamy problem. It is  possible that horizon fluctuations are small for large black holes if the appropriate semiclassical state is sharply peaked. But  if a firewall were to form, its accompanying back reaction on the metric would obviously lead to high curvature fluctuations, and in turn to large horizon fluctuations in the early stages of its formation. 

As a last comment, our quantization also demonstrates an exact duality between (a sector) of the model and 1+1 free scalar field theory on the half line. This  in turn is dual to fermionic theory via the well-known Bose-Fermi correspondence in two spacetime dimensions. This means that the quantization we have presented likely has a  fermionic description. 

This work is a  first exploration of the the use of dust time to study quantization of gravity. Natural extensions of our approach are to the 3+1 theory Bianchi models, spherically symmetry, and other  reduced sectors, such as the Gowdy models.

\begin{acknowledgments}

We thank Jorma Louko, Gabor Kunstatter and Sanjeev Seahra for helpful comments. This work was supported by NSERC of Canada, and an AARMS Postdoctoral Fellowship to J.Z.
\end{acknowledgments}

\bibliography{3dDust}

\end{document}